\documentclass[a4paper, amsfonts, amssymb, amsmath, reprint, showkeys, twoside, nofootinbib,superscriptaddress, floatfix]{revtex4-2}
\usepackage[english]{babel}
\usepackage[utf8]{inputenc}
\usepackage[colorinlistoftodos, color=green!40, prependcaption]{todonotes}
\usepackage{siunitx}
\usepackage{svg}
\usepackage{graphicx}
\usepackage[T1]{fontenc}
\usepackage[version=4]{mhchem}
\newcommand\thefont{\expandafter\string\the\font}
\usepackage{array}
\usepackage{amsmath}
\usepackage{mathtools}
\usepackage{mathrsfs}
\usepackage{mathptmx}
\usepackage{float}
\usepackage{braket}
\usepackage[hidelinks]{hyperref}
\usepackage{subcaption}
\usepackage{makecell}
\graphicspath{ {./Plots/} }
\DeclareGraphicsExtensions{.png, .pdf}

\begin{document} 

\title{Uncooled Thermal Infrared Detection Near the Fundamental Limit Using a Nanomechanical Resonator with a Broadband Absorber}
\author{P. Martini}
\affiliation{Institute of Sensor and Actuator Systems, TU Wien, Gusshausstrasse 27-29, 1040 Vienna, Austria.}
\author{S. Emminger}
\affiliation{Institute of Sensor and Actuator Systems, TU Wien, Gusshausstrasse 27-29, 1040 Vienna, Austria.}
\author{K. Kanellopulos}
\affiliation{Institute of Sensor and Actuator Systems, TU Wien, Gusshausstrasse 27-29, 1040 Vienna, Austria.}
\author{N. Luhmann}
\affiliation{Institute of Sensor and Actuator Systems, TU Wien, Gusshausstrasse 27-29, 1040 Vienna, Austria.}
\author{M. Piller}
\affiliation{Institute of Electronics, Graz University of Technology, 
Inffeldgasse 12/I, 8010 Graz, Austria.}
\author{R. G. West}
\affiliation{Institute of Sensor and Actuator Systems, TU Wien, Gusshausstrasse 27-29, 1040 Vienna, Austria.}
\author{S. Schmid}\email[Correspondence email address: ]{silvan.schmid@tuwien.ac.at}
\affiliation{Institute of Sensor and Actuator Systems, TU Wien, Gusshausstrasse 27-29, 1040 Vienna, Austria.}

\date{January 6, 2025}

\begin{abstract}
This paper introduces a thermal infrared detector utilizing a nano-optomechanical silicon nitride (SiN) resonator, equipped with a free-space impedance-matched (FSIM) absorber composed of a platinum (Pt) thin film, offering a broadband spectral absorptance on average of \SI{47}{}\%. To reduce photothermal back-action caused by intensity fluctuations of the readout laser, the FSIM absorber incorporates a circular clearance for the laser. The study provides a comprehensive characterization of the thermal time constant, power responsivity, and frequency stability of the resonators, with experimental results compared to analytical models and finite element method (FEM) simulations. The fastest thermal response is observed for the smallest \SI{1}{\milli\m} resonators, with a thermal time constant $\tau_{th}=$\SI{14}{\milli\second}. The noise equivalent power (NEP) of the resonators is assessed, showing that the smallest \SI{1}{\milli\m} resonators exhibit the best sensitivity, with NEP = \SI{27}{\pico\watt\per\sqrt\hertz} and a respective specific detectivity of D$^*$ = \SI{3.8e9}{\centi\meter\sqrt\hertz\per\watt}. This is less than three times below the theoretical maximum for an ideal IR detector with 50\% absorptance. This places our resonators among the most sensitive room-temperature IR detectors reported to date offering an extended spectral range from the near-IR to far-IR. This work underscores the potential of nano-optomechanical resonators for high-performance IR sensing applications.

\end{abstract}

\maketitle

\section{Introduction}
\label{sec:intro}

Infrared (IR) detectors are crucial tools in a wide range of applications, including spectroscopy, imaging, environmental monitoring, and thermal sensing. They can be broadly categorized into two types: photon detectors and thermal detectors \cite{rogalski2002infrared}.  Photon detectors, such as quantum well and HgCdTe (MCT) detectors, are highly sensitive but typically require cryogenic cooling to achieve optimal performance. In contrast, thermal IR detectors, such as thermopiles, bolometers, and pyroelectric detectors, operate effectively at room temperature, making them more practical and cost-effective for many applications. These thermal detectors measure IR radiation by converting it into a temperature change, which then induces a measurable response. This room-temperature operation, coupled with the ability to detect a broad spectral range, gives thermal IR detectors a distinct advantage in applications where cooling is impractical or where spectral broadband detection is required.

Despite advancements in uncooled thermal detectors, their sensitivity has remained significantly below the fundamental detection threshold at room temperature given by the specific detectivity D$^* = \SI{1.4e10}{\centi\meter\sqrt\hertz\per\watt}$ for front and backside coupled detectors, which is dictated by temperature fluctuations of both the detector and its surrounding environment \cite{rogalski2002infrared, bib:Datskos_detectors, kruse2004can, skidmore2003superconducting}. 
Conventional thermal detectors that rely on thermoelectric detection schemes are typically limited by electronic noise, e.g. Johnson noise or 1/f noise, which restrict their sensitivity, particularly when operating at room temperature. This noise originates from the electronic readout circuits that convert thermal signals into electrical signals, making it challenging to achieve a sensitivity close to the fundamental limit.

To overcome the existing limitations of thermoelectric detectors, alternative thermal detection mechanisms have been explored. One approach is thermal IR detection with a temperature-sensitive mechanical resonator.
The idea of a mechanical resonator used as thermal IR detectors dates back to 1969 \cite{cary1969infrared}. Microelectromechanical system-based (MEMS) IR detectors have been introduced in 1996 \cite{vig1996uncooled}. Advancement in nanofabrication techniques allowed the development of the first nanoelectromechanical system (NEMS) for frequency-shift-based infrared thermal sensing \cite{zhang2013nanomechanical}. This work was shortly followed by others \cite{yamada2013photothermal, hui2013high} and nowadays, NEMS resonators are used in applications ranging from IR spectroscopy \cite{luhmann2023nanoelectromechanical} to the detection of visible  \cite{blaikie2019fast}, IR \cite{bib:Piller_thermal}, and terahertz (THz) radiation \cite{zhang2024high,vicarelli2022micromechanical}.

To maximize the amount of absorbed IR light, a widely used strategy is to exploit the optical properties of metamaterials, such as plasmonic antennas, which reach almost 100\% absorptance \cite{das2023thermodynamically, wei2020metamaterial, cui2014plasmonic, ogawa2017wavelength, zhang2024high}. However, the exceptional absorption properties of such devices are limited to a narrow spectral range \cite{cui2014plasmonic, ogawa2017wavelength, zhang2024high} and their fabrication is complex \cite{das2023thermodynamically}.
Free space impedance-matched (FSIM) absorbers are an alternative to high-efficiency narrow-band absorbers. FSIM absorbers offer a nominal absorptance of 50\% over a wide spectral range spanning the entire infrared regime. A constant absorptance is essential, e.g., for use in IR spectroscopy. FSIM absorbers can be made of a few nanometer-thin metal films, such as Au \cite{luhmann2020ultrathin} or Pt \cite{bib:Piller_thermal}. The small thermal mass of such thin metal films is important to the detector's performance.

\begin{figure*}
    \centering
    \includegraphics[width=0.9\linewidth]{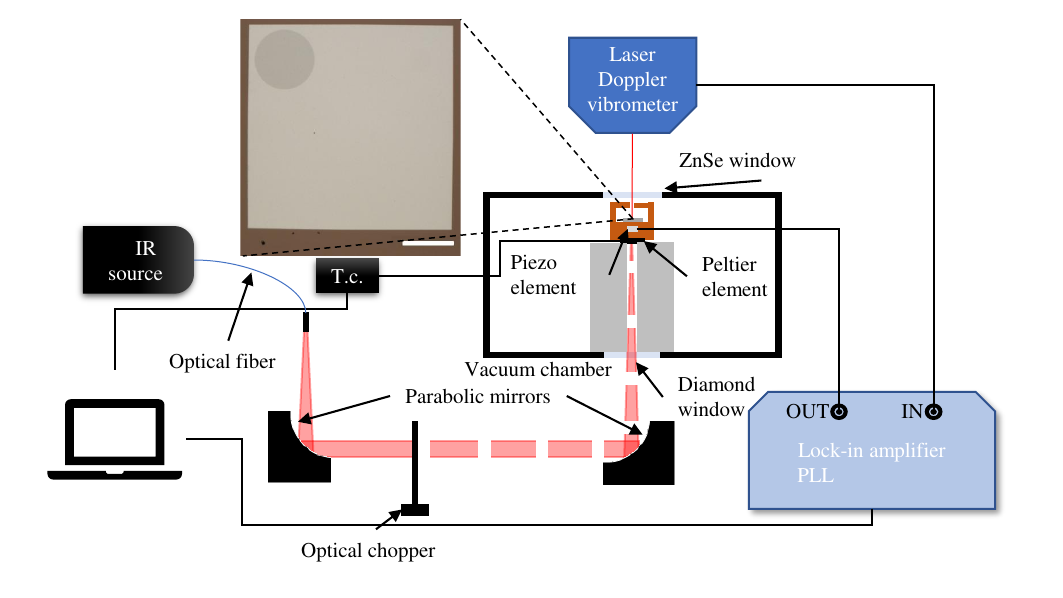}
    \caption{Schematic representation of the IR measurement setup. T.c. stands for temperature controller. Inset: optical microscope picture of a \SI{1}{\milli\m} membrane featuring an FSIM absorber with a visible circular clearance for the readout laser. The scale bar is \SI{200}{\micro\m}.}
    \label{fig:schematic}
\end{figure*}

While NEMS-based IR detectors featuring an FSIM absorber have achieved sensitivities as low as 7~pW/Hz$^{1/2}$ \cite{bib:Piller_thermal}, the sensitivity remained significantly below the fundamental limit. It has been shown that the readout noise of high-Q nanomechanical resonators can negatively affect the resonator's frequency stability, due to the so-called Leeson effect in oscillator circuits \cite{bevsic2023schemes, bib:Schmid_fundamentals,sadeghi2020frequency}.
Here, we present a nanomechanical IR detector with an optical interferometric readout instead of an electric readout. Such a square nanomechanical silicon nitride membrane resonator, featuring an FSIM absorber, is shown in Fig.~\ref{fig:schematic}. To mitigate frequency noise from the photothermal back action of the readout laser, which can be significant \cite{sadeghi2020frequency}, the membrane features a clearance in the FSIM absorber located at anti-nodal points of the (2,2) or (3,3) mode. Furthermore,
the detector is placed within an artificial thermal bath during measurements to stabilize the radiative heat transfer of the resonator with the environment.

The performance of IR detectors is characterized by the following figures of merit: the thermal time constant $\tau_{th}$, the noise equivalent power $NEP$, and the specific detectivity $D^*$ \cite{rogalski2002infrared}.
The NEP represents the sensitivity in units [\SI{}{\watt\per\sqrt\hertz}].
As nanomechanical resonators detect incident IR radiation via detuning of the resonance frequency, the NEP is given by  \cite{bib:Schmid_fundamentals}
\begin{equation}\label{eq:NEP}
    \text{NEP}(\omega) =\frac{\sqrt{S_y(\omega)}}{R_P(\omega)\alpha},
\end{equation}
where $S_y(\omega)$ is the frequency stability of the resonator (fractional frequency noise power spectral density) with units [\SI{}{1\per\hertz}], $R_P(\omega)$ is the relative power responsivity with units [\SI{}{1\per\watt}], and $\alpha$ is the absorptance of the NEMS detector. 

The responsivity of a nanomechanical resonator characterizes the detuning of its eigenfrequency $\omega_0$  per absorbed radiation power $P_0$ and is defined by \cite{bib:Schmid_fundamentals}
\begin{equation}\label{eq:R_P}
        R_P(\omega) = \frac{\partial \omega_0}{\partial P_0}\frac{1}{\omega_0} H_{th}(\omega)
\end{equation}
with the low-pass filter transfer function
\begin{equation}
    H_{th}(\omega)= \sqrt{\frac{1}{1 + (\omega \tau_{th})^2}}.
\end{equation}

For quantum detectors or thermal detectors operating at the quantum limit, the sensitivity scales with the square root of the detector area $A$ \cite{jones1953performance}. Normalizing the sensitivity with $A$ results in the 
specific detectivity D$^*$ in units [\SI{}{cm \sqrt\hertz\per\watt }]  \cite{bib:Datskos_detectors, nudelman1962detectivity}
\begin{equation}\label{eq:D*}
    \text{D}^* = \frac{\sqrt{A}}{\text{NEP}}.
\end{equation}
Subsequently, we evaluate the performance of the nanomechanical IR detector based on these figures of merit.

\section{\label{sec:methods} Methods}
\subsection{\label{subsec:fabrication}Resonator fabrication}
The resonators consist of \SI{50}{\nano\m} thick LPCVD deposited silicon nitride membranes of \SIlist{1000;2000;3000}{\micro\m} side lengths.  The initial nominal stress of the SiN was \SI{50}{\mega\pascal}. A Pt thin film, with a target thickness of \SI{5}{\nano\m}, was deposited by physical vapour deposition. The circular clearance was created by a lift-off process. Finally, the membranes were released from the backside by a KOH wet etch through the wafer.

\subsection{\label{sec:setup} Measurement setup}
For the measurements, a membrane is placed inside a custom-made, high-vacuum ($\sim\SI{e-5}{\milli\bar}$) chamber. This has the double benefit of increasing the Q factor (getting rid of medium losses \cite{bib:Schmid_fundamentals}) and reducing thermal losses (convection heat transfer can be neglected for pressures < $\SI{e-3}{\milli\bar}$). 
The chamber has two optical accesses, one for the measurement laser (top) and one for the IR light (bottom). The enclosures feature a silica glass window on the top and a diamond window (Diamond Materials GmbH) on the bottom, the latter specifically designed to ensure optimal transmission of mid-infrared (MIR) light.  Furthermore, the membrane is placed inside an artificial thermal bath made from a copper block. Both the lid and the bottom of the oven provide a hole aligned with the respective optical access of the chamber. A schematic representation of the measuring setup is displayed in Fig.~\ref{fig:schematic}. 

The IR source (Arclight-MIR from Arcoptix) is focused on the area of interest using an optical fiber (FIB-PIR-900-100 from Arcoptix) and two gold-coated, parabolic mirrors (from Thorlabs) with a focal length of \SI{101.6}{\milli\meter} and \SI{50.8}{\milli\meter}. An optical chopper can be placed between the mirrors to study the transient response of the resonator to characterize the thermal time constant of each membrane. 

The vibration of the resonator is detected using a laser Doppler vibrometer (LDV MSA-500 from Polytech). The equipment uses a \SI{633}{\nano\m} laser which is focused on the clearance in the absorber on the membrane. The reckoned signal (the frequency of vibration of the resonator) is fed into a lock-in amplifier (MFLI from Zurich Instruments). After pinpointing the frequency of the harmonic mode of interest, a closed loop is established with a phase-locked loop (PLL).

\section{Results \& Discussion}
\label{sec:results}

\subsection{Thermal time constant}
\begin{figure}
    \centering
    \includegraphics[width=\linewidth]{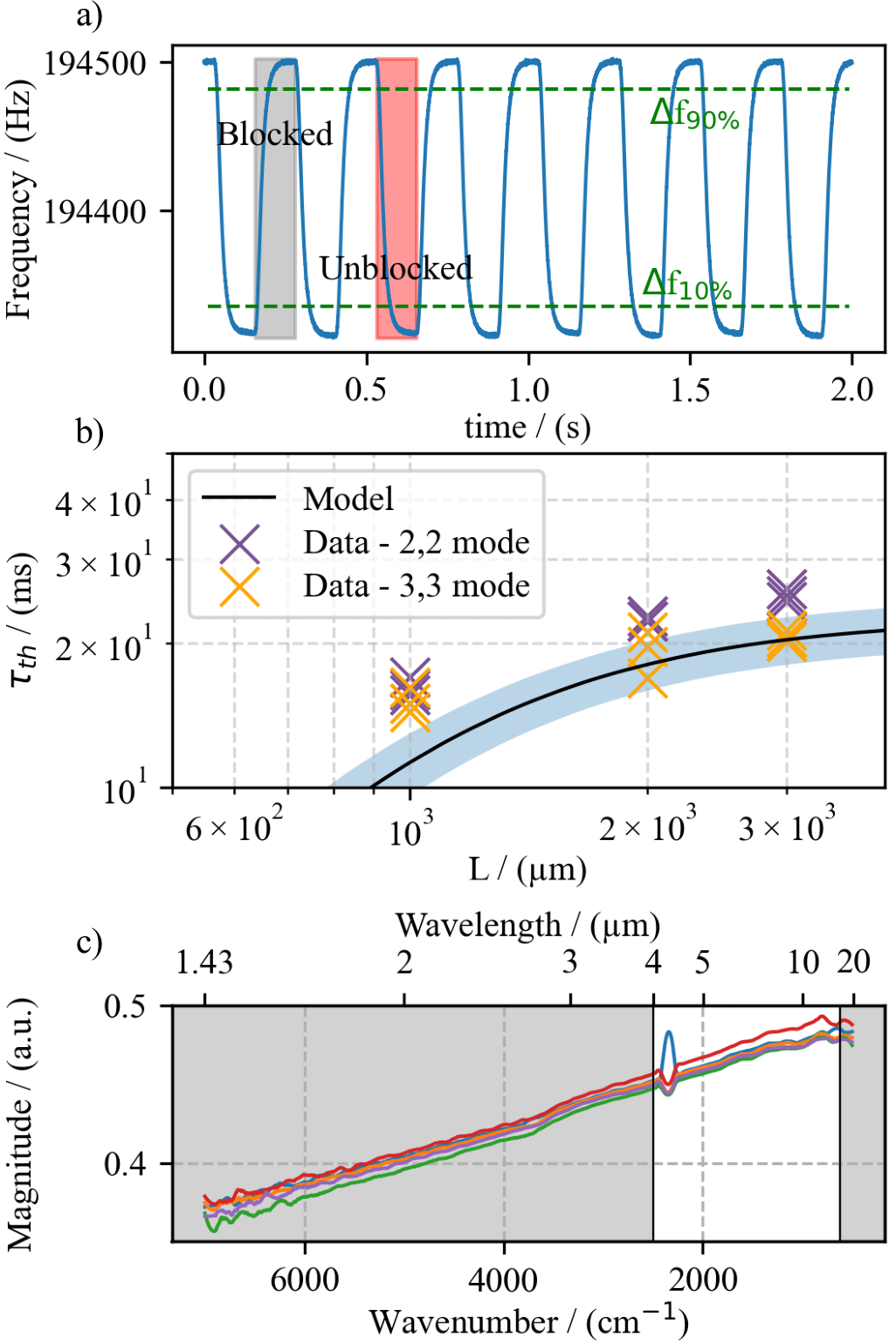}
    \caption{Thermal and optical behaviour of \SI{50}{\nano\meter} thick membranes covered with a thin Pt layer. a) Frequency signal of a \SI{1}{\milli\meter} membrane recorded when using the optical chopper. IR power is \SI{7.5}{\micro\watt}. $\Delta f_{90\%}$ and $\Delta f_{10\%}$ delimit the portion of the transient used for calculation of $\tau_{th}$. b) $\tau_{th}$ for \SI{18}{} membranes of different dimensions and comparison with the theoretical model. The blue area around the black curve represents the uncertainty band coming from different material parameters listed in Table \ref{tab:mat_par}. c) FTIR absorptance spectra of five \SI{1}{mm} membrane covered with a Pt thin-film-based FSIM absorber. Grey areas of the plots are the ranges not transmitted by the IR optical fiber.}
    \label{fig:tau}
\end{figure}

Fig.~\ref{fig:tau}a shows the frequency response to intensity-modulated IR radiation. The time constant was calculated using the \textit{90-10 method}.
The results of the measurements of the 18 membranes tested in this work are displayed in Fig.~\ref{fig:tau}b, together with the analytical model, given by \cite{bib:Schmid_fundamentals,bib:Piller_thermal,bib:kanellopulos_comparative}:
\begin{equation}\label{eq:tau_th_model}
    \tau_{th}=\frac{C}{G}
\end{equation}
with the heat capacity
\begin{equation}
    C = L^2 \sum_{i}h_{i}\rho_{i}c_{p_{i}}
\end{equation}
and  thermal conductance
\begin{equation}
    G = 8\pi \sum_{i}h_{i}\kappa_{i} + 8\epsilon\sigma_{SB}T^3,
\end{equation}
where, for every \textit{i}-th stacked material that composes the detector, $h$ is the thickness, $\rho$ the mass density, $c_{p}$ the specific heat capacity, and $\kappa$ is the thermal conductivity; $\epsilon$, $\sigma_{SB}$, and $T$ are the emissivity, the Stefan-Boltzmann constant, and the temperature, respectively.

The agreement with the theory (black lines) is good. The largest uncertainty in the model comes from the thermal conductivity ($\kappa$) of the ultrathin Pt film. In literature, different values can be found or calculated, ranging from \SI{29.5}{\watt\per\meter\kelvin} \cite{zhang2005thermal}  to \SI{14.5}{\watt\per\meter\kelvin} \cite{zhang2007thermal}. All material parameter-based uncertainty is represented in the plot through the colored blue band. The other material parameters for SiN and Pt used for the model are reported in Table~\ref{tab:mat_par}. 

Building on the findings of Luhmann et al. \cite{luhmann2020ultrathin}, based on the impedance-matched theory \cite{bib:Hilsum_IR}, the Pt thin film has a $50\%$ nominal absorbance over the entire mid-IR wavelength range. The Pt-based FSIM absorber, which typically provides a constant absorptance \cite{bib:Piller_thermal}, has a slight spectral dependency shown in Fig.~\ref{fig:tau}c. This is due to the Pt film thickness that is below the target thickness of 5~nm, as discussed in the Supplementary Information. Calculations of the effective emissivity yield a value of $\epsilon \approx 0.48$.

\begin{table*}[]
    \centering
    \begin{tabular}{ccccccc}
        \hline
         & \makecell{$E$\\$\SI{}{\giga\pascal}$} & \makecell{$\rho$\\$\SI{}{\kilo\gram\per\cubic\meter}$} & \makecell{$c_{p}$\\$\SI{}{\joule\per(\kilo\gram\kelvin)}$} & \makecell{$\kappa$\\$\SI{}{\watt\per(\meter\kelvin)}$} & \makecell{$\alpha_{th}$\\$\SI{}{\per\micro\kelvin}$} & \makecell{$\nu$\\} \\ 
        \hline \hline
        SiN & \SI{247\pm36}{} & \SI{3000 \pm 163}{} & \SI{775 \pm 56}{} & $3.2(0.5)$ & $1.6(0.5)$ & $0.27(0.03)$ \\
        
        Pt & \SI{143\pm25}{} & \SI{20742 \pm 630}{} & \SI{132.6 \pm 1.8}{} & $21.3(6.2)$ & $9.1(0.3)$ & $0.39(0.02)$ \\
   
        \hline
    \end{tabular}
    \caption{Material parameters for silicon nitride and platinum used in the analytical model. Given the uncertainty in the literature for such thin films, the mean values and standard deviations used in this work are provided.}
    \label{tab:mat_par}
\end{table*}

\subsection{Responsivity}
The calculation of NEP (\ref{eq:NEP}) requires the knowledge of the responsivity.
The responsivity (\ref{eq:R_P}) of a square membrane resonator for an even heating of the membrane can be derived analytically and yields \cite{bib:Schmid_fundamentals, bib:kanellopulos_comparative}
\begin{equation}\label{eq:resp_model}
    R_P(\omega) = \frac{\partial \omega_0}{\partial T}\frac{1}{\omega_0}\frac{\partial T}{\partial P}H_{th}(\omega) = \frac{R_T}{G}H_{th}(\omega)
\end{equation}
with the temperature responsivity
\begin{equation}
R_T = -\frac{\alpha_{th}}{2(1-\nu)}\frac{E}{\sigma}.
\end{equation}

\begin{figure*}
    \centering
    \includegraphics[width=\linewidth]{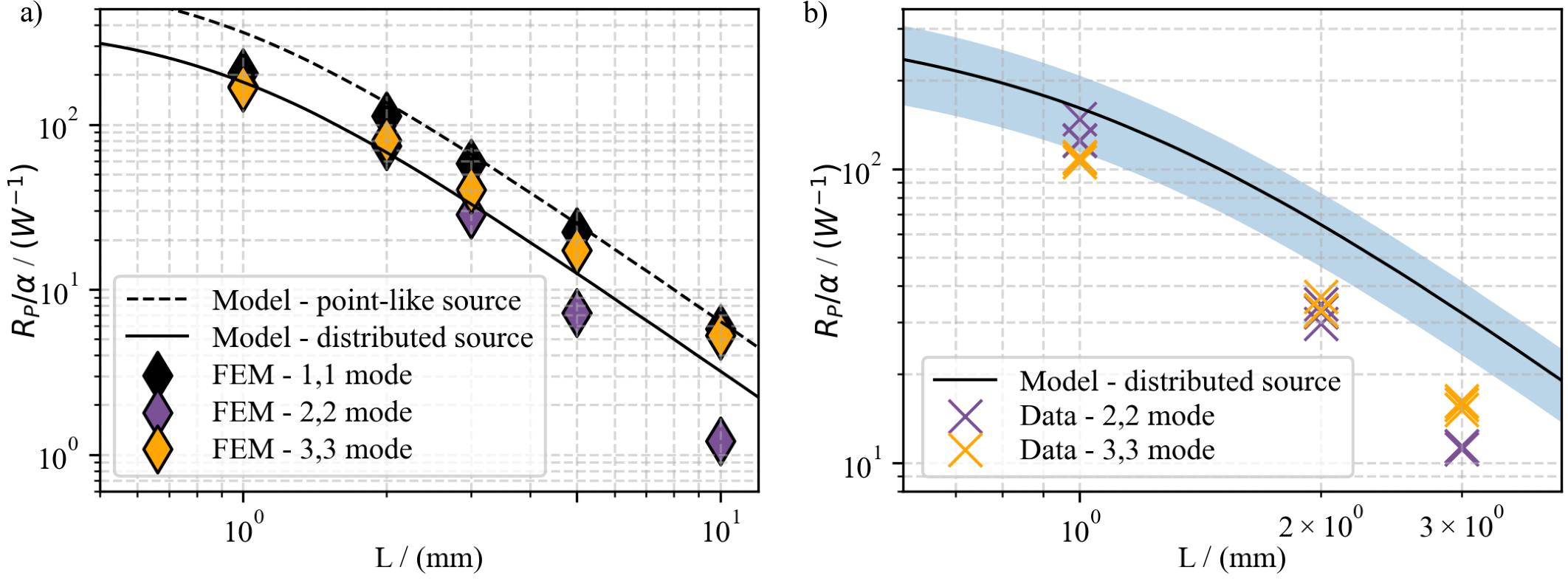}
    \caption{Steady-state power responsivity $R_P/\alpha$ for square membranes resonators. a) Comparison between the analytical models for a distributed source (solid line) and a point-like source (dashed line), alongside FEM simulations (COMSOL Multiphysics) for an IR beam with a \SI{600}{\micro\meter} diameter (rhombuses). b) Experimental measurements (crosses) compared to the distributed source model. The blue band indicates the model's uncertainty due to variations in material parameters (Table \ref{tab:mat_par}).}
    \label{fig:resp_comparison}
\end{figure*}

In Fig.~\ref{fig:resp_comparison}a absorptance-normalized responsivities computed with FEM for the given IR beam diameter of 600~$\mu$m are compared to the analytical model (\ref{eq:resp_model}) for an evenly distributed heating and for point-like heating. The latter model predicts values that are a factor of 2 larger \cite{bib:Schmid_fundamentals,bib:kanellopulos_comparative}. An absorptance of the Pt FSIM absorber over the spectral window of the IR light source effectively transmitted by the optical fiber is $\alpha=0.47$ (see Fig.~\ref{fig:tau}c). The comparison shows that the model for even heating is the correct model approximation for the smallest membranes, while the point-like model is the correct approximation for the larger membranes.

The comparison of the analytical model to FEM simulations further reveals that the model accurately predicts the (1,1) and (3,3) modes, while it shows a significant deviation for the (2,2) mode. 
This discrepancy for the (2,2) mode can be explained by the strong mismatch between the temperature field and the displacement field. The former has a maximum while the latter has a minimum in the membrane center. The overlap between the two fields is better for uneven modes than it is for even modes. A more detailed investigation of the relationship between mode shape responsivity is presented in the Supplementary Information.

Fig.~\ref{fig:resp_comparison}b compares the measured responsivities to (\ref{eq:resp_model}). The IR light power impinging on the membranes is $P = \SI{7.5}{\micro\watt}$. The measured values for the smallest membranes match the model within the given uncertainties. The responsivities for the larger membranes drop below the model predictions, which is expected for higher order modes, in particular the (2,2) mode, as predicted from the FEM data (Fig.~\ref{fig:resp_comparison}a).

\subsection{Frequency stability}\label{subsec:noise}
In addition to the responsivity, NEP (\ref{eq:NEP}) is defined by the frequency stability. 
The main sources of frequency noise in our nanomechanical photothermal system come from \cite{bib:kanellopulos_comparative}: additive phase noise $S_{y_{\theta}}(\omega)$, temperature fluctuation noise $S_{y_{th}}(\omega)$, and photothermal back-action noise $S_{y,\delta P}(\omega)$. 

$S_{y,\delta P}(\omega)$ is the frequency noise induced by power fluctuations of the readout laser (photothermal back-action). It readily can be modeled as \cite{bib:kanellopulos_comparative}:
\begin{equation}\label{eq:photothermal_backaction_nosie_PSD}
    S_{y_{\delta P}}(\omega) = \alpha^2 R_{P}^{2} S_{I}(\omega),
\end{equation}
where $S_{I}(\omega)$, with units [W$^2$/Hz], is the source's relative intensity fluctuation power spectral density. For a typical laser, this is the sum of three components \cite{maddaloni2013laser}: \textit{i)} laser shot noise, \textit{ii)} flicker noise, and \textit{iii)} random walk noise.

Frequency stability is best analyzed through a frequency signal's Allan variance, which can be calculated by
\begin{equation}
    \sigma_y^2(\tau) =\frac{1}{2\pi}\frac{8}{\tau^2}\int_0^\infty\frac{\left[\sin(\omega\tau/2) \right]^4}{\omega^2}S_y(\omega)\text{d}\omega.
\end{equation}

\begin{figure}
    \centering
    \includegraphics[width=\linewidth]{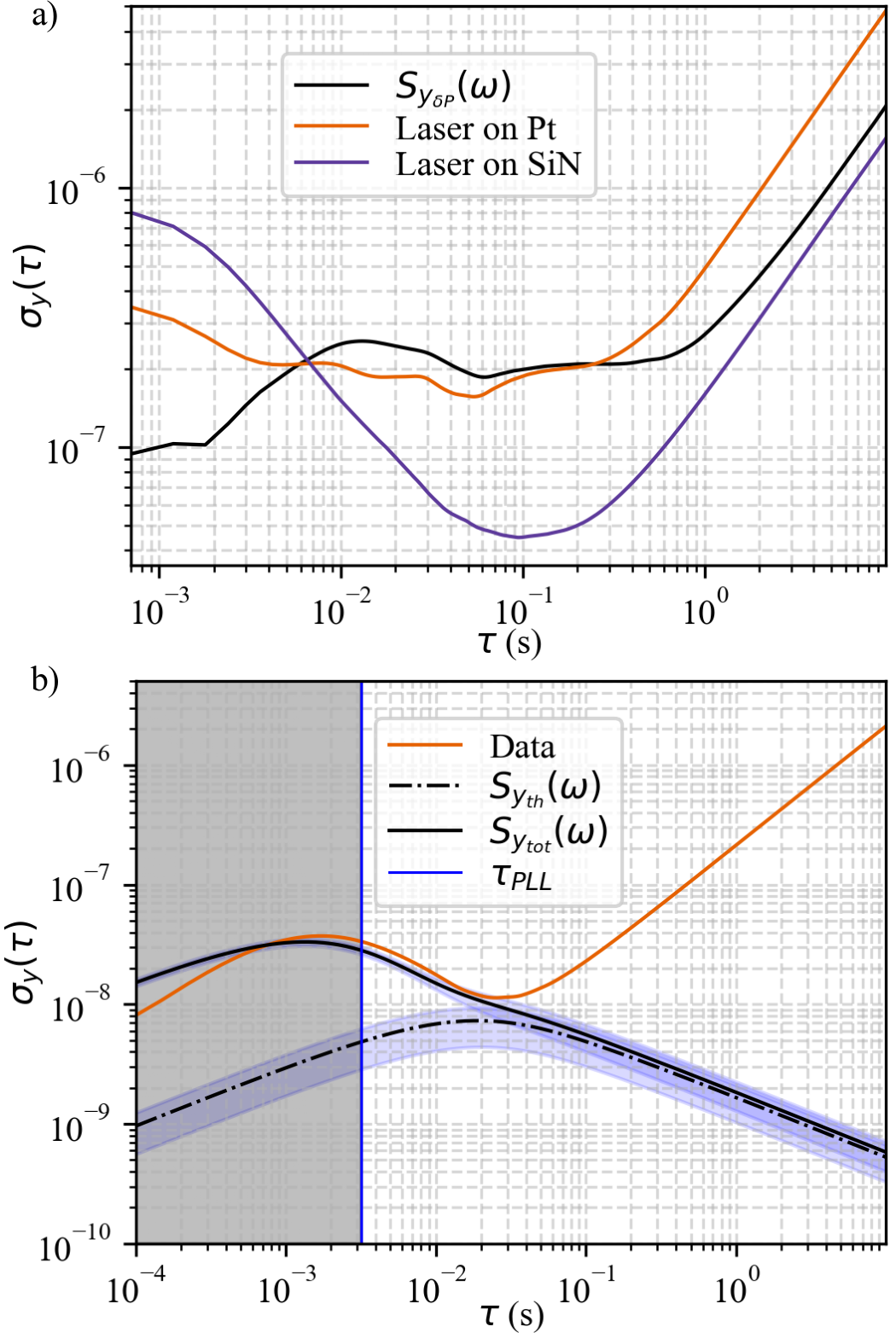}
    \caption{Frequency stability study by means of Allan deviations. a) Study of the photothermal back-action frequency noise induced by the readout laser for a \SI{1}{\milli\meter} membrane. The laser power was \SI{15.4}{\micro\watt}. b) Frequency stability of the (2,2) mode of a well-performing \SI{2}{\milli\meter} resonator compared to theoretical thermal fluctuation noise (dotted curve) and total phase noise (solid black curve). The experimental detection to thermomechanical noise ratio was $\mathcal{K} = \SI{0.006}{}$, the laser power was \SI{6.78}{\micro\watt} and the thermomechanical peak amplitude was $S_{y_{th}}(\omega_0) = $~\SI{9.8e-23}{\meter\squared\per\hertz}}
    \label{fig:AD_Pt_vs_SiN}
\end{figure}

Photothermal back-action noise has been shown to limit the frequency stability in silicon nitride string resonators \cite{sadeghi2020frequency}. For a given laser intensity noise, (\ref{eq:photothermal_backaction_nosie_PSD}) can be minimized by reducing the absorptance $\alpha$. The solution we implemented in this work is to spare an area in the Pt absorber thin film (see Fig.~\ref{fig:schematic}) so that the readout laser can be reflected off the silicon nitride instead of the Pt absorber.
The effectiveness of this solution is demonstrated in Fig.~\ref{fig:AD_Pt_vs_SiN}a, which compares the frequency stability (in terms of the Allan deviation) for the two scenarios. On the one hand, the frequency stability strongly deteriorates when the readout laser is pointed directly at the Pt absorber. On the other hand, when pointed at the clearance, the frequency stability is unaffected by photothermal back-action. The measured Allan deviation obtained on the Pt matches the calculated Allan deviation (\ref{eq:photothermal_backaction_nosie_PSD}) based on the measured laser's intensity fluctuation spectral density $S_I(\omega)$. This data shows the effectiveness of pointing the readout laser onto the clearance in the Pt absorber, by which photothermal back-action can be circumvented.

Additive phase noise $S_{y_{\theta}}$ is the result of the conversion of thermomechanical noise $S_{z_{thm}}(\omega)$ and detection noise $S_{z_{d}}(\omega)$ into phase noise, ultimately resulting into frequency noise \cite{bevsic2023schemes,demir2021understanding}:
\begin{equation}\label{eq:thermomechanical_noise_PSD}
            S_{y_{\theta}}(\omega) = \frac{1}{2 Q^2}\frac{S_{z_{thm}}(\omega_0)}{z_0^2}\left[\lvert H_{\theta_{thm}}(\mathrm{i}\omega)\lvert^2 + \mathcal{K}^2 \lvert H_{\theta_{d}}(\mathrm{i}\omega)\lvert^2 \right],    
\end{equation}
with the resonator's quality factor $Q$, eigenfrequency $\omega_0$,  displacement amplitude $z_0$, and thermomechanical amplitude noise at resonance
\begin{equation}
    S_{z_{thm}}(\omega_0)=\frac{4 k_B T Q}{m_{eff} \omega_0^3},
\end{equation}
with the Boltzmann constant $k_B$, temperature $T$, and effective resonator mass $m_{eff}$.
Detection noise is defined relative to the resonator's thermomechanical noise peak as follows $S_{z_{d}}(\omega)= \mathcal{K}^2 S_{z_{thm}}(\omega_0)$. Finally, the two transfer functions in (\ref{eq:thermomechanical_noise_PSD}) for a phase-locked loop are given by \cite{bevsic2023schemes}
\begin{equation}
\begin{split}
    H_{\theta_{thm}}(\mathrm{i}\omega) &= \frac{H_L(\omega)}{H_L(\omega)+\mathrm{i}\omega\tau_{PLL}}\\
    H_{\theta_{d}}(\mathrm{i}\omega) &= \frac{1}{H_R(\omega)}H_{\theta_{thm}}(\omega) 
\end{split}
\end{equation}
with the low-pass transfer functions $H_R(\omega)$ and $H_L(\omega)$ of the resonator, with a time constant $\tau_R=2Q/\omega_0$, and the oscillator circuit, respectively.

Temperature fluctuation noise $S_{y_{th}}(\omega)$ defines the ultimate sensitivity limit of a thermal detector \cite{bib:Schmid_fundamentals}. It originates from the heat exchange of a thermal detector with the environment, both via conduction and radiation. Because of the statistical nature of this process, temperature fluctuation noise can be assumed white. These random fluctuations provoke a shift in the resonance frequency that depends on the resonator's thermal conductance $G$ and temperature responsivity $R_{T}$. This fractional frequency-noise PSD for an optomechanical sensor can be modelled with \cite{bib:Schmid_fundamentals,bib:kanellopulos_comparative}: 
\begin{equation}\label{eq:temp_fluc_noise_PSD}
            S_{y_{th}}(\omega) = \frac{4 \kappa_{B} T^2}{G_{eff}} R^{2}_{T} H_{th}^2(\omega).
\end{equation}
with the effective thermal conductance
\begin{equation}\label{eq:G_eff}
    G_{eff} = 4\pi \sum_{i}h_{i}\kappa_{i} + 8 L^2 \varepsilon \sigma_{SB} T^3
\end{equation}
with the Boltzmann's constant $\kappa_{B}$.

Fig.~\ref{fig:AD_Pt_vs_SiN}b shows a comparison of an experimental to the theoretical frequency stability of the best-performing membrane resonator. The measured Allan deviations closely follow the modelled values dominated by additive phase noise, that is, thermomechanical and detection noise. For longer integration times, the frequency stability would approach the ultimate limit given by temperature fluctuation frequency noise. For this specific membrane the thermal time constant is $\tau_{th}=\SI{22.3}{\milli\second}$.

\subsection{\label{subsec:NEP_results}NEP and D$^*$}

\begin{figure*}
    \centering
    \includegraphics[width=\linewidth]{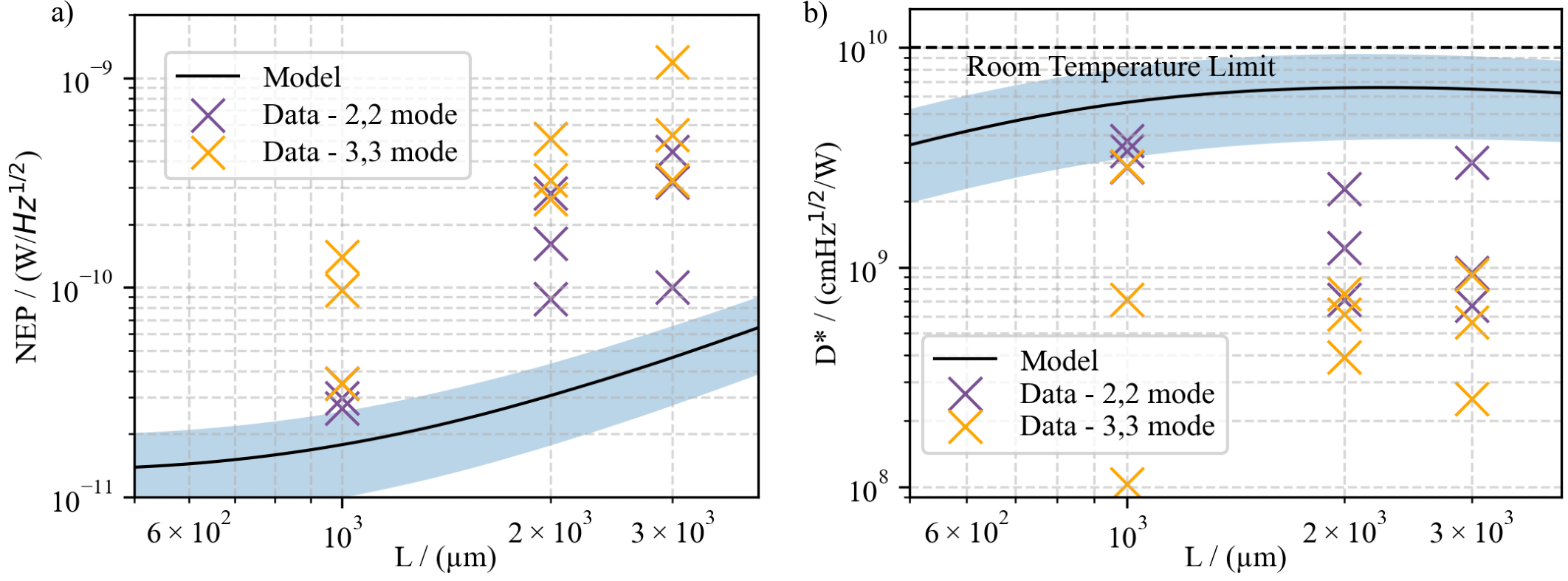}
    \caption{The noise equivalent power (NEP) and specific detectivity (D*) of membranes optimized for the second and third harmonic modes. a) NEP values were calculated using the power spectral density (PSD) at the thermal time constant frequency; b) The D* was also calculated. The results indicate minimal differences in performance between the different modes and dimensions of the membranes, with values approaching the fundamental limit at room temperature calculated for $50\%$ absorptance.}
    \label{fig:NEPandD*}
\end{figure*}

Fig.~\ref{fig:NEPandD*} presents the measured NEP and D$^*$ values compared to theory. The performance varies strongly between different resonators, with the best values of NEP = \SI{27}{\pico\watt\per\sqrt\hertz} and D$^*$ = \SI{3.8e9}{\centi\meter\sqrt\hertz\per\watt} for a 1~mm resonator in the (2,2) mode, approaching the theoretically predicted limit due to temperature fluctuations, which can be obtained by assuming  a perfectly isolated detector from (right term in (\ref{eq:G_eff})) with  (\ref{eq:temp_fluc_noise_PSD}) and (\ref{eq:D*}), which yields \cite{bib:Schmid_fundamentals}
\begin{equation}\label{eq:D-limit}
    \textbf{D}^* = \sqrt{\frac{\varepsilon}{32\sigma_{SB}k_BT^5}}.
\end{equation}

The ultimate specific detectivity for a nominal emissivity $\varepsilon=0.5$ then becomes D$^*\approx$\SI{1.0e10}{\centi\meter\sqrt\hertz\per\watt} at room temperature.
The measured value is only a factor of 3 below this fundamental limit indicated by the horizontal dashed black line in Fig.~\ref{fig:NEPandD*}b. For an IR detector with a perfect absorber ($\varepsilon = 1$), the specific detectivity limit would increase only slightly to  D$^*\approx$\SI{1.4e10}{\centi\meter\sqrt\hertz\per\watt}. The plot further reveals that the theoretical performance of a membrane resonator remains slightly below the fundamental limit.

\section{\label{sec:conclusions}Conclusions}

The results for NEP and D$^*$ demonstrate that our sensors are among the most sensitive room-temperature IR detectors. The minimum noise equivalent power achieved is \SI{27}{\pico\watt\per\sqrt\hertz}, and the maximum specific detectivity reaches \SI{3.8e9}{\centi\meter\sqrt\hertz\per\watt}. The latter value is less than a factor of 3 below the fundamental limit at room temperature for an IR detector with 50\% absorptance. The obtained sensitivity is on par with state-of-the-art uncooled optomechanical infrared detectors featuring subwavelength-structured meta-absorbers \cite{zhang2024high,das2023thermodynamically}. In contrast to previous approaches, the FSIM absorber extends the spectral range from the near-infrared to the terahertz regime. Simultaneously, its nominal 50\% absorptance results in only a 1.4-fold reduction in ultimate sensitivity. As a result, the achieved sensitivity is less than a factor of 4 below the fundamental limit for an ideal IR detector with 100\% absorptance.

The performance of the presented nanomechanical IR detector is on par with state-of-the-art commercial pyroelectric detectors with a NEP and specific detectivity of \SI{40}{\pico\watt\per\sqrt\hertz} and \SI{4e9}{\centi\meter\sqrt\hertz\per\watt}, respectively, at 10~Hz.\footnote{LT31 Single Channel Voltage Mode DLaTGS Pyroelectric Detectors from Laser Components} 
As demonstrated in this work, the nanomechanical membrane detectors achieve the expected theoretical performance, albeit below the fundamental limit. To further approach the thermal fluctuation noise limit, we aim to explore trampoline resonators as a promising alternative, leveraging their superior thermomechanical properties \cite{bib:kanellopulos_comparative}. Combined with the FSIM absorber strategy employed here, trampoline resonators hold the potential to advance room-temperature IR detectors to the fundamental sensitivity threshold.

\section{Acknowledgments}
The author would like to thank Tatjana Penn and Veljko Vukicevic for their help in the design of the vacuum chamber and assembling the measurement setup. Thanks to Michael Buchholz for the help with the metal deposition.

\newpage
\bibliographystyle{unsrt}
\bibliography{bib}

\end{document}


\title{Supplementary Information: Uncooled Thermal Infrared Detection Near the Fundamental Limit Using a Nanomechanical Resonator with a Broadband Absorber}

\author{P. Martini}
\affiliation{Institute of Sensor and Actuator Systems, TU Wien, Gusshausstrasse 27-29, 1040 Vienna, Austria.}
\author{S. Emminger}
\affiliation{Institute of Sensor and Actuator Systems, TU Wien, Gusshausstrasse 27-29, 1040 Vienna, Austria.}
\author{K. Kanellopulos}
\affiliation{Institute of Sensor and Actuator Systems, TU Wien, Gusshausstrasse 27-29, 1040 Vienna, Austria.}
\author{N. Luhmann}
\affiliation{Institute of Sensor and Actuator Systems, TU Wien, Gusshausstrasse 27-29, 1040 Vienna, Austria.}
\author{M. Piller}
\affiliation{Institute of Electronics, Graz University of Technology, 
Inffeldgasse 12/I, 8010 Graz, Austria}
\author{R. G. West}
\affiliation{Institute of Sensor and Actuator Systems, TU Wien, Gusshausstrasse 27-29, 1040 Vienna, Austria.}
\author{S. Schmid}\email[Correspondence email address: ]{silvan.schmid@tuwien.ac.at}
\affiliation{Institute of Sensor and Actuator Systems, TU Wien, Gusshausstrasse 27-29, 1040 Vienna, Austria.}

\maketitle

\date{\today}

\section{Calculation of Platinum absorptance}
\begin{figure}[b]
    \centering
    \includegraphics[width=\linewidth]{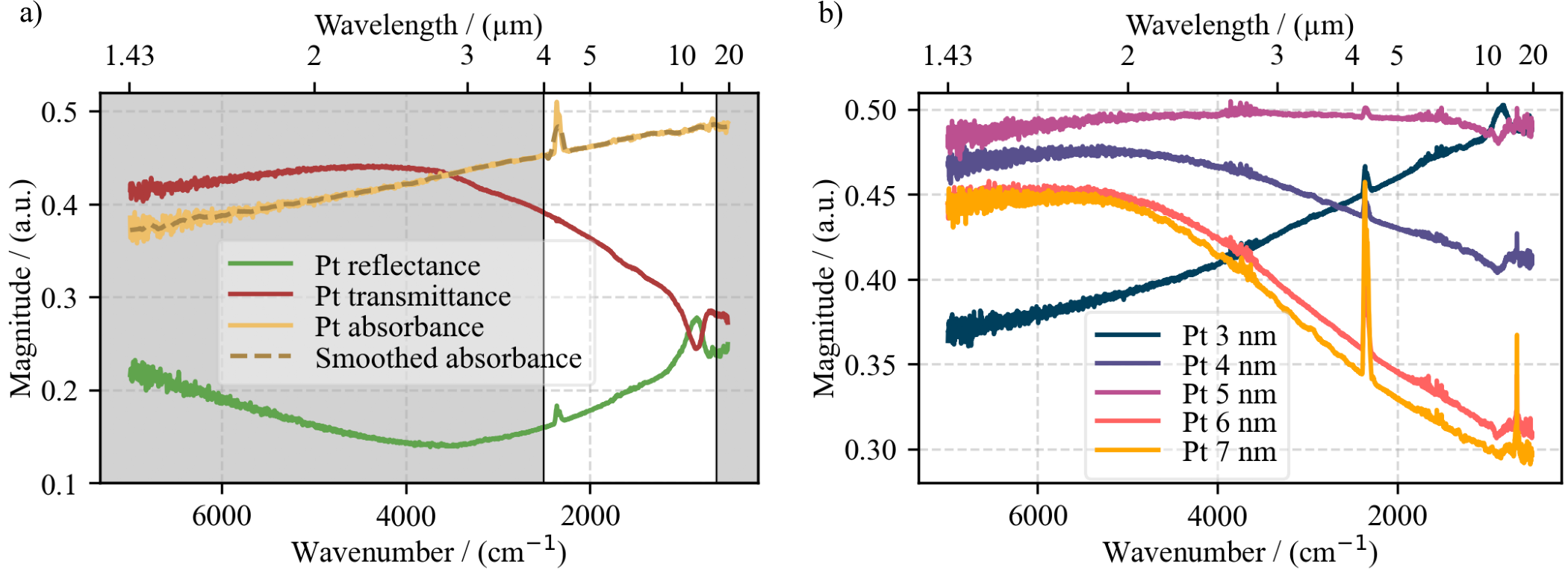}
    \caption{FTIR spectra of \SI{1}{mm} membrane covered with a thin Pt layer. a) Reflectance (green line) and transmittance (red line) spectra used to calculate the absorptance spectrum (yellow line); the dashed line is the result of the smoothening operation. The ranges not transmitted by the IR optical fiber are grey areas of the plots. b) Absorptance spectra for different thicknesses of Pt thin layers over SiN.}
    \label{fig:Pt_abs_spectra}
\end{figure}

For the comparison between the measured responsivity value and the analytical model, the absorptance of the Pt layer $\alpha$ is required. To determine this value, we measured five $\SI{1}{mm}$ drums using an FTIR spectrometer (Tensor 27 FT-IR spectrometer from Bruker). These membranes were from the same batch as those used for the IR measurements. The absorptance spectra are calculated from the transmittance and reflectance spectra obtained. Figure \ref{fig:Pt_abs_spectra}a shows a representative FTIR measurement result. The grey-shaded areas on the plot represent regions of the spectrum not transmitted by the IR optical fiber. Therefore the value of $\alpha$ is calculated by averaging the absorptance curve only over the spectrum region that is effectively transmitted (white area of the plot). A spectral feature is observed at a wavenumber of \SI{2565}{\per\cm}. This is a characteristic and well-known feature of \ce{CO2}. In the present case, this feature comes from the presence of \ce{CO2} in the FTIR measuring chamber because of the operator's breathing while loading the chip. To mitigate this feature and reduce the noise in the spectra at the edges of the FTIR range, we applied a \textit{Savitzky-Golay} filter to the absorptance spectrum. The resulting smoothed curve is shown as the dashed line in Figure \ref{fig:Pt_abs_spectra}a.

Comparing these FTIR absorptance measurements with those for different thicknesses of Pt on SiN reveals that the Pt layer thickness deposited in this work has an approximate thickness of \SI{3}{\nm} (dark blue line in Figure \ref{fig:Pt_abs_spectra}b). The targeted thickness was \SI{5}{\nm}, at which the Pt exhibits a flat absorptance of $\SI{0.5}{}$ (purple line in Figure \ref{fig:Pt_abs_spectra}b) across a wide range of IR wavelengths. This represents the maximum achievable absorptivity in the IR for a thin metal film, as reported by Hilsum \cite{bib:Hilsum_IR}. Nevertheless, when calculating the average of the absorptance curve over the wavelength range transmitted by the IR fiber and averaging across all five drums, we obtain a value of $\alpha=\SI{0.47}{}$, not far from the optimal value.

\newpage
\section{Power responsivity and modeshape relationship}

\begin{figure}[b]
    \centering
    \includegraphics[width=\linewidth]{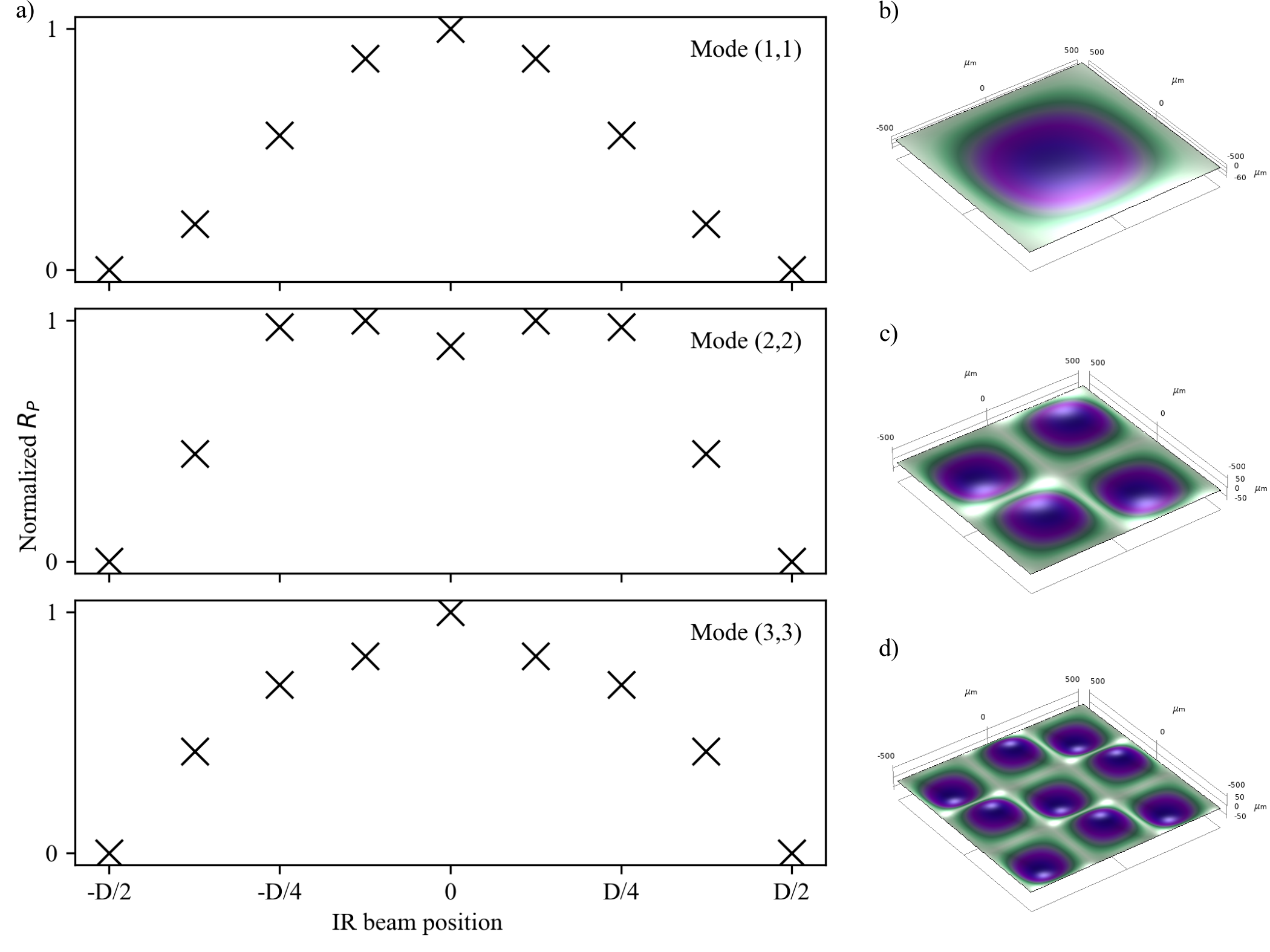}
    \caption{Normalized power responsivity and modeshape relations. a) Normalized responsivity results from FEM simulations when scanning a point-like source over 9 points across the diagonal (D) of a square membrane. b), c) and d) FEM modeshape representation of the (1,1), (2,2) and (3,3) modes, respectively.}
    \label{fig:modeshape_responsivity}
\end{figure}

The normalized power responsivity for a \SI{1}{\mm} membrane is displayed in Figure \ref{fig:modeshape_responsivity}a, for the first three harmonic modes. Here, the simulated power source has a diameter of \SI{40}{\nm} (point-like source), and it is scanned over nine different points across one diagonal of the drum. The zero on the x-axis represents the center of the membrane, while the first and last point in each graph refers to two opposite vertices of the square, which is the resonator. 

The COMSOL modeshape representation for the first three harmonic modes is displayed in Figures \ref{fig:modeshape_responsivity}b, \ref{fig:modeshape_responsivity}c and \ref{fig:modeshape_responsivity}d. For the (3,3) mode, the point of maximum responsivity is located in the center, where there is the maximum displacement, and at the same time, it is the point with the best thermal isolation (it is the point furthest away from the edge). The same is true for the fundamental (1,1) mode, and, by extension, for every vibrational mode with an antinode in correspondence with the center. For the (2,2) mode, the point with the best thermal isolation does not correspond to the point with maximum deflection. In this case, the best responsivity performance is achieved between the antinode of the modeshape and the center of the membrane. This explains why the match between the measured responsivity of the (3,3) and the fundamental mode and the values predicted by the analytical model is better than the results for the (2,2) mode.

\clearpage
\bibliographystyle{unsrt}
\bibliography{bib}